# Neutrino Telescopes in the Mediterranean Sea

**Juan José Hernández-Rey**[1]

IFIC – Instituto de Física Corpuscular, CSIC-Universitat de València, apdo. 22085, E-46017 Valencia, Spain

Juan.J.Hernandez@ific.uv.es

**Abstract**. The observation of high energy extraterrestrial neutrinos can be an invaluable source of information about the most energetic phenomena in the Universe. Neutrinos can shed light on the processes that accelerate charge particles in an incredibly wide range of energies both within and outside our Galaxy. They can also help to investigate the nature of the dark matter that pervades the Universe. The unique properties of the neutrino make it peerless as a cosmic messenger, enabling the study of dense and distant astrophysical objects at high energy. The experimental challenge, however, is enormous. Due to the weakly interacting nature of neutrinos and the expected low fluxes very large detectors are required.

In this paper we briefly review the neutrino telescopes under the Mediterranean Sea that are operating or in progress. The first line of the ANTARES telescope started to take data in March 2006 and the full 12-line detector was completed in May 2008. By January 2009 more than one thousand neutrino events had been reconstructed. Some of the results of ANTARES will be reviewed. The NESTOR and NEMO projects have made a lot of progress to demonstrate the feasibility of their proposed technological solutions. Finally, the project of a $km^3$-scale telescope, KM3NeT, is rapidly progressing: a conceptual design report was published in 2008 and a technical design report is expected to be delivered by the end of 2009.

## 1. Introduction

Being weakly interacting, stable and neutral, the neutrino is a unique cosmic messenger that can provide a good deal of physics information. Astrophysical neutrinos point back to their source, can come from the cores of dense objects and travel long distances without interacting.

Astrophysical objects that are able to accelerate protons and nuclei in a wide energy range are known to exist and it is likely that some of these accelerated cosmic rays produce neutrinos when interacting with matter or radiation either at the medium surrounding their source or in their travel to us. Active galactic nuclei (AGN) and gamma-ray bursts (GRB) have been proposed as possible cosmic accelerators outside the Galaxy, whilst supernova remnants (SNR) are the main candidates to accelerate cosmic rays within our Galaxy. Other galactic astrophysical objects such as pulsar wind nebulae and micro-quasars have been put forward as possible neutrino sources. Recently, some observations in x- and gamma rays point to hadronic acceleration mechanisms in the production of some of the observed high energy gamma rays. Neutrinos could help to clarify the physics behind the high energy astrophysical phenomena and to elucidate the sources of high energy cosmic rays.

---

[1] on behalf of the ANTARES Collaboration and the KM3NeT Consortium.

The main challenge for Neutrino Astronomy is the huge size required to record a sufficient number of events given the low expected fluxes. A suitable technique is the detection of neutrinos by the observation of the Cherenkov light induced by their interaction products when traversing natural media such the ice of the South Pole or the water of deep oceans or lakes. These media act both as a shield for downgoing muons coming from normal cosmic rays showers produced in the atmosphere and as a dark, optically transparent material suitable to observe the Cherenkov radiation of the neutrino products. An array of photo-sensors at sufficiently large depth can record the Cherenkov light induced by upgoing muon tracks that can only come from neutrinos that have traversed the Earth and interacted relatively close to the array. Thanks to the long muon path length the effective volume for this channel is especially high, but other neutrino flavours can also be detected through the hadronic and electromagnetic showers they produce. The muon channel, moreover, provides a very high angular resolution since at high energies the angle between the muon and its parent neutrino is low and the resolution is mainly limited by instrumental effects (optical properties of the medium and time and spatial resolution of the telescope). The energy of the neutrino can be estimated within a factor 2-3 through the energy deposited in the detector by the corresponding track or shower.

Sea or lake water and ice are therefore large and cheap natural media than can be used to deploy a huge neutrino detector using the Earth as a natural shield. The main backgrounds for this sort of telescope are downgoing atmospheric muons that could be wrongly reconstructed as upgoing and the unavoidable flux of upgoing muons coming from atmospheric neutrinos produced by cosmic ray showers at the other side of the Earth. The challenge is therefore to detect an excess of neutrinos either diffuse (in all directions) or point-like (localized in a given point of the sky) above this irreducible background.

The present status of neutrino telescopes worldwide is as follows [1]. The AMANDA telescope, deployed in the ice of the South Pole, has been taking data since more than a decade and its successor, IceCube, is at present operating with 59 of its total 80 planned strings and it will be completed in two years. In the northern hemisphere, the BAIKAL detector is operating since more than a decade. In the Mediterranean Sea, the ANTARES telescope is taking data since 2006 in a partial configuration and since 2008 in its full set-up. The NESTOR and NEMO projects have made considerable progress in the testing of their proposed technologies. A large undersea infrastructure, KM3NeT, that will host a $km^3$ telescope is under design and its construction is expected to start by 2011.

In this paper we review the different neutrino telescope projects in the Mediterranean Sea, paying special attention to the results of the ANTARES telescope and the progress of the KM3NeT initiative.

## 2. The ANTARES telescope

The ANTARES detector [2] is the first large neutrino telescope in the deep sea. It is located 40 km off Toulon in the French coast (42º 50' N, 6º 10' E) at a depth of 2500 m. The telescope consists of 12 lines anchored to the seabed and held vertically taut by buoys. Each line is 450 m long and consists of 25 floors separated by a vertical distance of 14.5 m, with the lowest floor located at 100 m above the seabed. Each floor contains three optical modules (OMs) [3] separated 120º horizontally from each other and looking 45º vertically downwards to optimize the detection of Cherenkov photons coming from upgoing tracks.

An OM consists of a pressure resistant glass sphere that contains and protects a 10" Hamamatsu photomultiplier (PMT). In each storey there is a local control module (LCM) which consists of a titanium container that houses the control and readout electronics together with a compass and a tiltmeter used for the reconstruction of the line's shape and rotation. A set of five storeys, called a sector, is controlled by a special LCM named the master LCM.

There is one hydrophone and one LED optical beacon (LOB) every sector. The hydrophones are part of an acoustic positioning system that is also composed of emitters at the bottom of the lines and in small pyramids located at the seabed and surrounding the telescope. The lines are flexible and can move under the influence of the sea current; the position of their components needs to be known for a proper track reconstruction. For typical water currents of 5 cm/s, the larger displacement –that takes place at the top of the line- is of the order of a few meters. The data of the acoustic positioning system are combined with those of the tiltmeters and compasses to obtain the shape of the lines through an overall fit to a formula that describes the shape as a function of the sea current. At present, precisions of the order of 10 cm are reached with the positioning system

The LOBs, five along each line, are part of an in-situ timing calibration system that also comprises two Laser Beacons (LBs) located at the bottom of the two central lines [4]. The computation of the time offsets between OMs is first carried out on shore during the assembly of the lines with a system of lasers and optical fibres. A 20 MHz clock signal generated on shore travels through the detector and ensures a precision of a fraction of nanosecond up to the LCM where a clock card is located. This precision is obtained thanks to an in-built echo system that returns the clock signal and allows the computation of the return time. The delays between the LCM (clock card) and the PMT are monitored and re-measured by means of the optical beacon system. The relative time offsets between OMs are measured with a precision of the order of 0.5 ns. The corrections of the time offsets with respect to those measured on shore are lower than 1 ns in 85% of the cases. These relative offsets show a high stability with time, therefore time calibration runs are taken only every month. The time offsets measured by the LOBs are in good agreement with those extracted from $^{40}$K coincidences in pairs of OMs in the same storey. The optical beacon system also shows that the time spread introduced by the readout electronics is within specifications (~0.2 ns).

For any photomultiplier charge signal exceeding that which would correspond to 1/3 of a photoelectron at nominal gain, its digitized time and amplitude are transmitted to shore. The data are processed in a computer farm where events with interesting features are written to disk and the rest discarded [5].

ANTARES started to take data with its first full line in March 2006 and by September 2007 a detector with 5 lines was already operative. The last two of the twelve lines were connected in May 2008.

2.1. First results

Good data runs recorded with the 5-line configuration were selected on the basis of a high number of active channels (>80%) and low bioluminescence (<40% of active channels in burst state). The selected run sample amounts to a live time of 168 days[2].

Two independent reconstruction algorithms and selection sets were used to identify upgoing muon tracks. The results of both methods are in agreement with expectations from the Monte Carlo simulation. In Figure 1, the distribution of the selected number of events as a function of the sine of the elevation angle (angle above the horizon) of the reconstructed track, sinθ, is shown for one of the reconstruction and selection strategies. In the left plot, the logarithm of the number of events as a function of sinθ is shown for all angles. Positive values (resp. negative) correspond to down-going (resp. upgoing) tracks. The experimental points are represented in black (statistical errors are negligible for downgoing tracks). The expectation from simulation for atmospheric muons (red histogram) and atmospheric neutrinos (blue histogram) or both (violet) are also shown. The band (in brown colour) represents the uncertainty in the Monte Carlo prediction. The number of upgoing selected events is 168 (one per day of live time), in agreement with the prediction from Monte Carlo simulation, 164 ± 3(stat) ± 33(theor) ±16(syst), where "theor" indicates the error stemming from the uncertainty in the theoretical predictions of the atmospheric neutrino flux and "syst" is the systematic error coming from the uncertainty in the detector simulation.

---

[2] A problem in the trigger software caused some data loss during event filtering. The actual equivalent live time of the final run sample for a perfectly working trigger would have been 140 days.

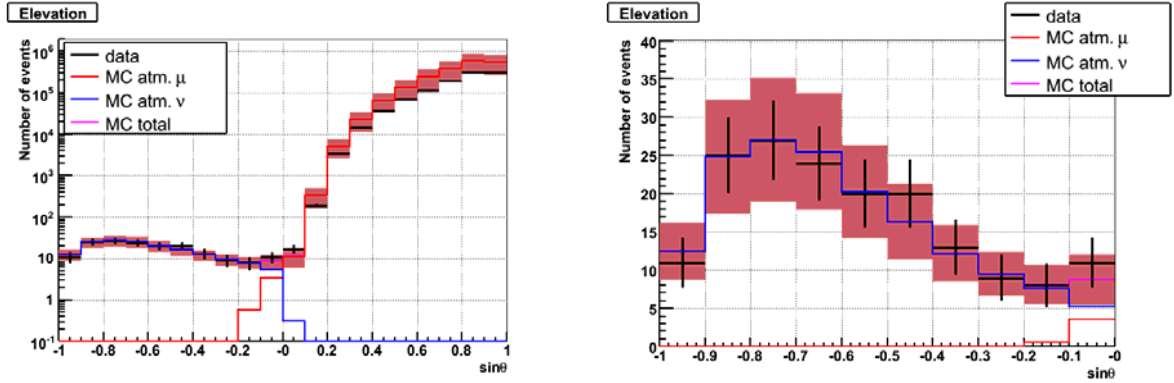

Figure 1 : Distribution of elevation angle of the selected events reconstructed using the data taken by the ANTARES 5-line detector. Left: distribution of the logarithm of the number of events as a function of the sine of the elevation angle. Right: distribution of the number of events for upgoing tracks.

This data set was used to look for possible steady neutrino point-like sources. A more stringent reconstruction and selection strategy was used in order to ensure high quality tracks and a good angular resolution. A total of 94 upgoing events were selected with this strategy. Before applying any search algorithm, the time of the events was scrambled providing realistic data samples for background estimation. A list of 25 possible astrophysical sources spread along the declination range observable by ANTARES was made up based on a variety of criteria (astrophysical objects detected by high energy gamma telescopes, the centre of the Galaxy, a source close to high energy cosmic rays observed by the Auger observatory, etc). A binned method based on cones of optimized size around the sources and an unbinned method based on the expectation-maximization algorithm [6] were used to look for an excess of events coming from any of the selected sources. After "data unblinding" and application of both methods, no clear signal was found. From the compatibility of the results with the background, the corresponding limits for a neutrino flux coming from the sources were set.

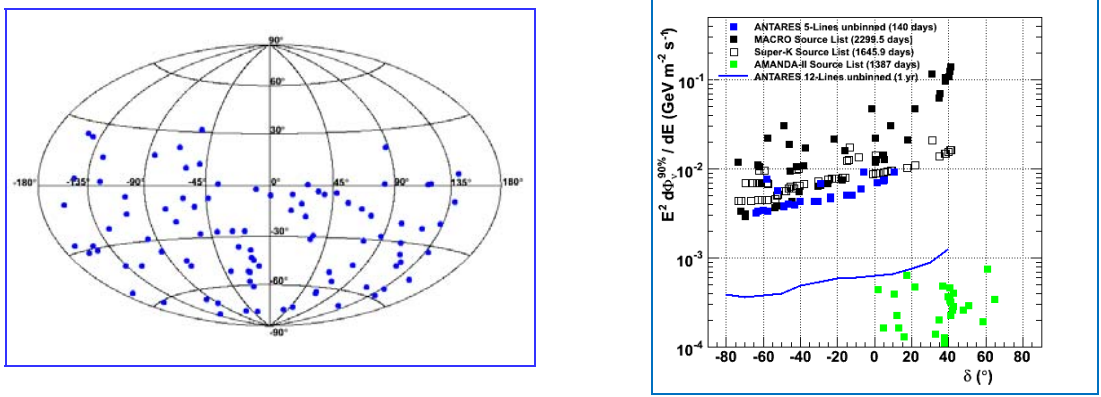

Figure 2: Left, distribution in equatorial coordinates of the 94 selected events for the steady point sources analysis. Right, the blue squares give the 90% C.L. limits on a neutrino flux, $E^2 \, d\Phi/dE$, as a function of the declination of the 25 selected astrophysical sources. The blue line gives the expected limit for the full ANTARES detector after one year of data taking.

In Figure 2 Left, the position in equatorial coordinates of the 94 selected neutrino events is given. In the right plot, the blue squares give the 90% C.L. limits on $E^2 \, d\Phi/dE$ for a neutrino flux, $\Phi$, coming from the 25 selected sources as a function of their declination. The white, black and green squares are the limits set by MACRO [7], SuperKamiokande [8] for the southern sky and AMANDA-II [9] for the northern sky, respectively. The blue line is the sensitivity expected for the full ANTARES telescope after one year of data taking.

The detector has been taking data with different configuration (9, 10 and 12 lines) since December 2007. As of January 2009 more than one thousand neutrino events have been reconstructed. Analysis of the data is proceeding well and results from a variety of physics topics, such as the search for neutrino diffuse fluxes and point sources, neutrino events in coincidence with GRBs, search for monopoles and nuclearites, etc, are expected soon.

## 3. The NESTOR and NEMO projects

The first initiative for a neutrino telescope in the Mediterranean Sea was the NESTOR project that started in 1989. The site is located in the Ionian Sea close to Pylos in the Peloponese at a depth of about 4000 m. The basic unit of the telescope is a tower composed of 12 stars stacked vertically with a total length of 410 m from the seabed. Each star has 6 titanium arms, 16 m long, holding at its tips two OMs housing 13" PMTs [10]. A prototype consisting of a reduced star with 5 m long arms with 12 OMs was deployed in 2003. More than 5 million triggers of 4-fold coincidences were recorded and a measurement of the atmospheric muon flux and zenith distribution was made [10]. The NESTOR collaboration is at present preparing a tower of five floors which together with four autonomous strings will make the NuBE arrangement (Neutrino Burst Experiment) to search for neutrinos in coincidence with gamma-ray bursts [10].

The NEMO collaboration started in 1998. The basic unit of their telescope concept is a semi-rigid tower composed of 16 storeys separated 40 m vertically. Each storey is composed of a 20 m rigid bar holding at its tip four OMs. The final selected site is close to Capo Passero in the south-east of Sicily (36° 19' N, 16° 5' E) around 80 km from the coast with an average depth of 3500 m. The Phase 1 of NEMO was intended to test prototypes of the main elements of the detector. A junction box and a mini-tower were deployed in December 2006 in a test site at 2100 m depth 25 km off Catania [11]. The mini-tower lost buoyancy after some time, but 200 hours of data were taken and downgoing muons were reconstructed [11]. In Phase 2 of NEMO a fully equipped 16 storey tower will be deployed at the final site in Capo Passero, the 100 km long cable linking the shore station with the site was laid in July 2007.

## 4. The KM3NeT project

The low expected fluxes of extra-terrestrial neutrinos call for a large telescope of a km$^3$ scale. In February 2006 a Design Study funded by the VI$^{th}$ Framework Program of the European Commission for a km$^3$-scale telescope in the Mediterranean Sea −dubbed KM3NeT− was initiated. This design study will conclude by the end of 2009 with the production of a Technical Design Report. In April 2008 a Conceptual Design Report providing the key concepts and the general parameters of the telescope was delivered [12]. Meanwhile, a Preparatory Phase, funded by the VII$^{th}$ Framework Program of the European Commission, meant to address the organizational, political and financial aspects of KM3NeT was initiated in 2008 and is planned to end with the start of the telescope construction by the end of 2011. KM3NeT will be a large multidisciplinary infrastructure [13], it will not only host the neutrino telescope but also a large variety of equipment that will provide a platform for research in a variety of earth and sea science disciplines.

KM3NeT was selected in 2005 by the European Strategic Forum for Research Infrastructures (ESFRI), together with other 35 research infrastructures in all fields of science, as being mature enough to proceed to its design in Europe and later confirmed in the 2006 ESFRI report [14]. KM3NeT is also one the seven infrastructures in Astroparticle Physics selected in the ASPERA ERA-Net Roadmap [15] and included in the ASTRONET ERA-Net Roadmap [16].

4.1. Detector design

The KM3NeT design requirements include an instrumented volume of at least 1 km$^3$, an angular resolution of about 0.1º for neutrino energies above 10 TeV, an energy threshold of a few hundreds of GeV and the capability to detect all neutrino flavours.

A variety of sea floor and 3D layouts have been studied in simulation (see Figure 3). Different concepts for the detector unit −the basic deployable detection structure that is individually anchored to the seabed− have been also investigated. Among these, we can mention the extended tower structure (rigid star-like storeys), the compact tower structure (compact, transportable horizontal bars) and cable based structures based on rolled up lines with either multiple or single optical modules arranged on each storey.

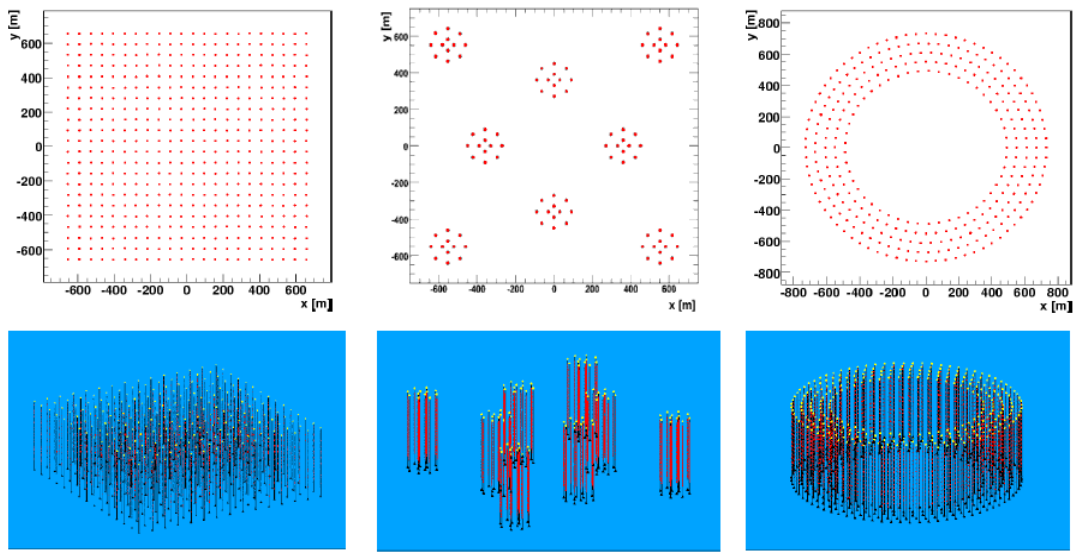

Figure 3: Layouts in the seafloor (top) and in 3D (bottom) of possible neutrino configurations with detector units arranged (from left to right) in homogeneous, clustered and ring geometries.

Several solutions have been proposed for the photo-sensor to be used inside the OMs. In addition to the classical large photocathode hemispherical photomultiplier (PMT), a direction-sensitive optical module based on a segmented PMT plus a mirror system has been put forward. Likewise, an OM based on many (~30) small PMTs and an (hybrid) X-HPD based optical module have also been proposed. Each solution has its advantages and drawbacks and the final selection will have to take into account different variables including complexity, reliability and cost, in addition to the technical merits of each proposal.

Independently of any final decision, a "reference detector" was selected in the KM3NeT CDR in order to perform simulation studies on the telescope sensitivity and physics reach. This reference detector consists of a homogeneous configuration with 225 (15x15) detection units each carrying 37 storeys with one OM composed of 21 3-inch PMTs covering its lower hemisphere. The distance between the detector units is 95 m and adjacent storeys are separated vertically by 15.5 m.

In Figure 4 left, the estimated limit for a diffuse flux as a function of the energy is shown for the KM3NeT reference detector together with the limits from AMANDA, IceCube and ANTARES (expected). In the right plot, the estimated limit for point sources as a function of declination for the reference detector is shown together with the measured or expected limits from other telescopes both for the northern and the southern sky.

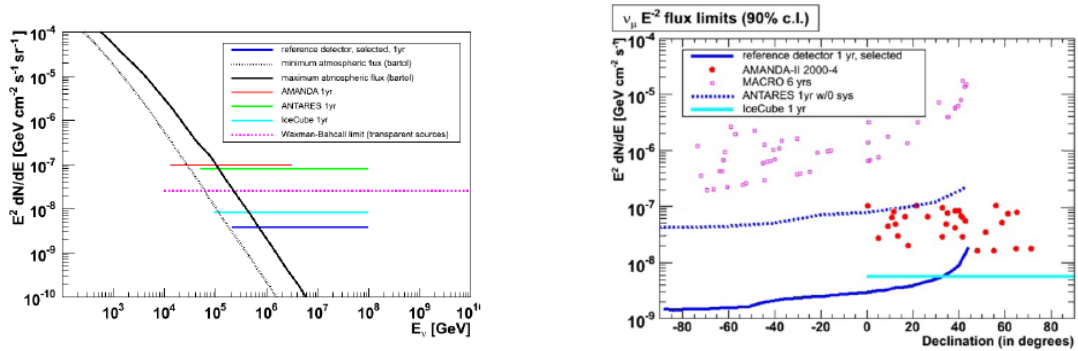

Figure 4: Left, comparison of the estimated diffuse flux limits for the KM3NeT reference detector to the different results from experiments and model fluxes. Right, point source sensitivity of the reference detector compared with the experimental limits and expected sensitivities of several telescopes.

The KM3NeT design study is performing other studies in a large variety of fields. KM3NeT will be in fact a multidisciplinary infrastructure that will provide a platform for research in a variety of earth and sea science disciplines such as seismology, gravimetry, radioactivity, geomagnetism, oceanography, geology and geochemistry.

## 5. Conclusions

Tremendous progress has been made during the last years in the field of Neutrino Astronomy. The ANTARES telescope has recently joined the already operating neutrino telescopes in the ice of the South Pole and in Lake Baikal. ANTARES has now set the more stringent experimental limits on neutrino point-like sources in the southern sky. Other initiatives such as NEMO and NESTOR made great technological progress towards the construction of a km$^3$-scale detector under the sea. The ANTARES, NEMO and NESTOR collaborations have joined their efforts in the KM3NeT project that aims to build such a telescope in the Mediterranean Sea. The KM3NeT consortium has already delivered a conceptual design report and is due to provide a technical design report by the end of 2009.

**Acknowledgments**
The support of the Spanish Ministerio de Ciencia e Innovación (grants FPA2006-04277, FPA2008-00961-E, CAC-2007-12, ICTS-2008-02) and of the VI$^{th}$ and VII$^{th}$ Framework Programmes of the European Commission (grant agreements 011937 and 212525) is gratefully acknowledged.